\documentclass{PoS}
\usepackage{graphicx}
\usepackage{epsfig,graphics,rotate}
\usepackage{amssymb}
\usepackage{amsmath}
\usepackage{amsfonts}
\usepackage{lipsum}

\title{Constraining Dark Matter Neutrino Interaction with High Energy Neutrinos}
\ShortTitle{}
\author{Carlos Arg\"uelles $^a$, \speaker{Ali Kheirandish} $^b$, and Aaron C. Vincent $^c$ \\
\llap{$^a$} Dept. of Physics, Massachusetts Institute of Technology, Cambridge, MA 02139, USA, \\ \llap{$^b$} Dept. of Physics and Wisconsin IceCube Particle Astrophysics Center, University of Wisconsin, Madison, WI 53706, USA, \\ \llap{$^c$} Department of Physics, Imperial College London, London SW7 2AZ, UK \\ 
\email{caad@mit.edu},\, \email{akheirandish@icecube.wisc.edu},\, \email{aaron.vincent@imperial.ac.uk}  \\}

\abstract{We employ data from the recently observed high-energy neutrino events at the IceCube Neutrino Observatory to constrain interactions between the dark matter (DM) in the Milky Way and the neutrino sector. We construct an extended un-binned  likelihood in order to explore the parameter space of allowed interactions. We present results in the specific case of a scalar DM candidate interacting via a scalar mediator, and show that due to the energy dependence of the interaction cross section, this approach can constrain the coupling more strongly than traditional cosmological probes for some regions of the parameter space. }
\FullConference{38th International Conference on High Energy Physics\\
  3-10 August 2016\\
  Chicago, USA}
\begin{document}
\section{Introduction}\label{Intro}
The search for dark matter (DM) interactions with the standard model (SM) sector has traditionally focused on weak-scale interactions with quarks, as these are the more readily observable, e.g. via elastic scattering in underground direct detection experiments. However, interactions between the DM and other particles may also be present and could be the dominant, or even sole, link between the dark and visible sectors. A DM-neutrino interaction is especially attractive for light DM models, where annihilation into heavier products is kinematically forbidden, and appears naturally in some models, for example when the DM is the sterile neutrino (see \cite{Adhikari:2016bei}).  DM-neutrino elastic scattering has been studied extensively in the literature, almost exclusively in the context of early universe cosmology 
(see \cite{Escudero:2015yka} and references therein). The transfer of power from DM to neutrinos before recombination can suppress power on small scales in the cosmic microwave background (CMB) and large scale structure. These effects take place at low ($\sim$ eV) temperatures. In particle physics, cross sections typically show some energy dependence $\propto E^2$ up to some scale. It therefore makes sense to turn to higher energies to probe such effects. Here, we search for the effect of DM-neutrino scattering of high-energy (TeV--PeV) neutrinos as they traverse the Milky Way halo. Ref. \cite{Davis:2015rza} considered a similar effect, though they did not precisely model the scattering cross section at high energies, nor did they compare their expected signal with experimental data. 

After four years of data taking, IceCube has confirmed observation of 53 High Energy Starting Events (HESE) with  greater than $6\sigma$ of being extraterrestrial origin \cite{Aartsen:2015zva}, rejecting purely atmospheric explanation. The arrival direction of these events is consistent with the isotropic distribution. This isotropy has been used to constrain a galactic contribution to the observed neutrino flux; either from standard sources \cite{Ahlers:2015moa} or from the decay or annihilation of halo DM \cite{Bai:2013nga}.

In this work, we present a novel approach to constrain the strength of the DM-neutrino interaction by using high energy cosmic neutrinos observed by IceCube. The DM-neutrino interaction strength would influence the isotropy of the extragalactic signal. As they pass through the galaxy on their way to earth, the flux of interacting neutrinos would be preferentially attenuated in the direction of the galactic centre, where the DM column density is the largest. For large enough coupling strengths, this should lead to an observable anisotropy in the neutrino sky.

In these proceedings we illustrate this effect with a single simplified model of DM-neutrino interaction, which is meant to represent the phenomenological behavior of a UV-complete model of the SM + dark sector at the energy scales of interest. We choose a scalar DM candidate, coupled to the neutrino sector via a scalar mediator. A full exploration of the set of possible models will be done in an upcoming work \cite{CAAinprep}. 

\section{Dark Matter-Neutrino Interactions}\label{models}
\begin{figure}[t!]
\center
\includegraphics[height=0.27\textheight]{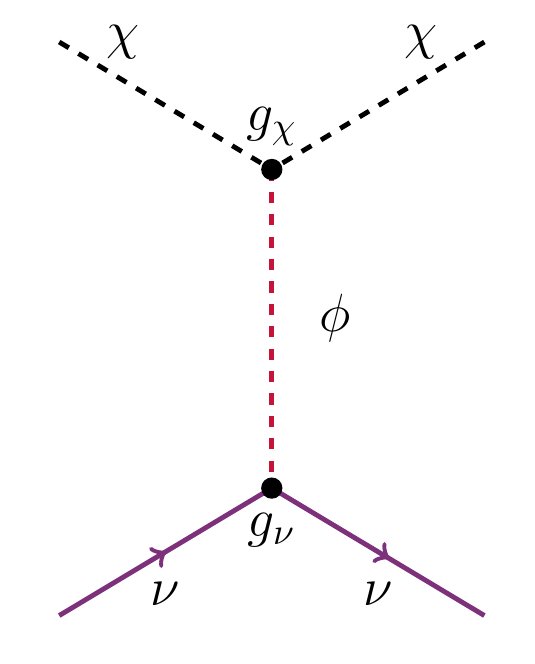}\includegraphics[height=0.27\textheight]{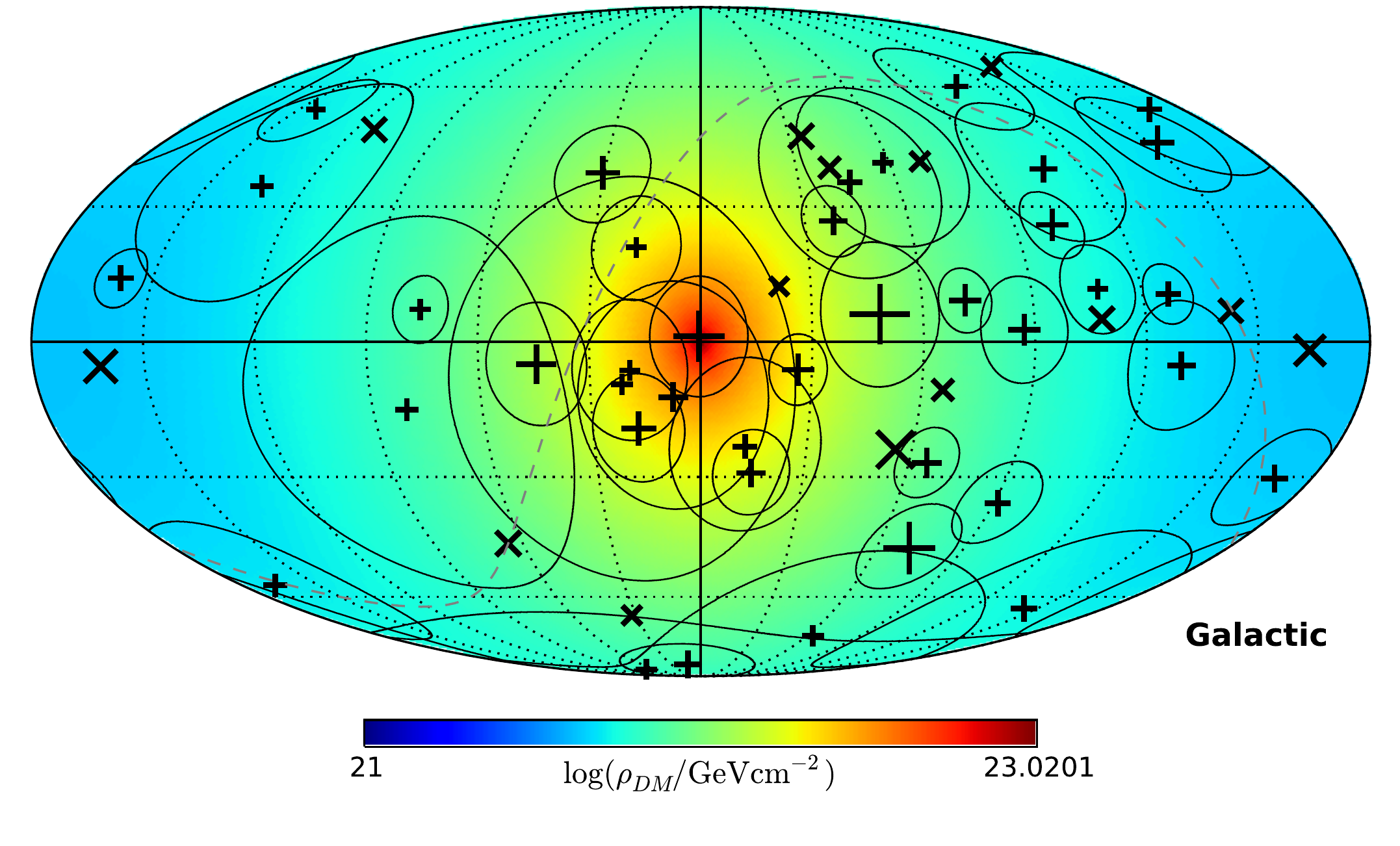}
\caption{Left: the simplified DM-neutrino interaction model considered here. Right: the arrival direction of the high energy starting events observed in 4 years of IceCube data \cite{Aartsen:2015zva} in galactic coordinates. Crosses represent shower events, while x's correspond to tracks. Circles represent the median angular uncertainty of cascades. The color scale is the column density of DM traversed by neutrinos arriving from each direction.}
\label{fig:skymap}
\end{figure}

The existence of a new mediator connecting the dark sector to neutrinos leads to a non-vanishing elastic scattering cross section. In this work, we focus on a single simplified model which give rise to potentially observable neutrino-DM interactions in dark matter halo: a scalar DM particle with a scalar mediator. We will present results for a more complete set of models in an upcoming publication \cite{CAAinprep}. This model is shown in the left-hand panel of Fig.~\ref{fig:skymap}: we call the DM particle $\chi$ and the mediator $\phi$. The coupling $g_\nu$ between the neutrinos and the mediator is similar to a Yukawa coupling, and dimensionless, whereas $g_\chi$ has dimensions of energy and can be assumed to encode higher-energy physical effects which have been integrated out. 

We compute the differential and total scattering cross sections, $\sigma$, in the frame of the Galaxy, where the DM is at rest, since non-relativistic thermal velocities can safely be neglected.

We take the incoming neutrino flux to be isotropic, assuming extragalactic in origin, and model the spectrum, $\phi(E)$, as a simple $E^{-2}$ power law, the expected behavior from Fermi acceleration mechanisms. As the neutrinos propagate towards the Earth, they must traverse the diffuse DM halo of the Milky Way, and in particular through the very DM dense Galactic center. Each arrival direction is therefore subject to a different column density of DM, and thus a different scattering rate, which is reflected as an anisotropic attenuation of the signal observed at Earth.

The attenuation of the extragalactic high-energy neutrino flux is described by the following cascade equation
\begin{equation} \label{eqcascade}
\frac{d\phi(E)}{d\tau} = -\sigma(E)\phi(E) + \int d\tilde E \frac{d\sigma(\tilde E,E)}{d E}\phi(\tilde E), 
\end{equation}
where $\tau$ is the DM column density and $E$ is the neutrino energy.
The first term on the right-hand side of Eq. \eqref{eqcascade} accounts for the down-scattering of neutrinos from energy $E$ to any other, while the second term accounts for the reverse effect: scattering of neutrinos from any other energy $\tilde E$ to $E$.

To model DM distribution of the Milky Way, we take an Einasto profile with shape parameters that fit the Via Lactea II simulation results \cite{2009Sci...325..970K}, and a local DM density of 0.3 GeV cm$^{-3}$. The DM column density and the arrival direction of high-energy cosmic neutrinos are shown in Fig. \ref{fig:skymap}.

\section{Analysis}\label{ana}
The observed neutrino flux is predominantly extragalactic, compatible with an isotropic distribution, and has flavor composition compatible with $(\nu_e:\nu_\mu:\nu_\tau) = (1:1:1)$ which follows the expected composition from a mainly-pionic origin of neutrinos and robust in the presence of new physics \cite{Arguelles:2015dca}.
Nonetheless, a different flavor composition at production will yield an oscillation-averaged flux that is very close to $(1:1:1)$, and with current statistics, would not be distinguishable within the space of flavors allowed by oscillation~\cite{Vincent:2016nut}. 
	
Based on these assumptions, we construct the following un-binned likelihood function for a given parameter set $\vartheta$ and set of observed topologies, energies, and arrival directions $\{t,E, \vec x\}$ 
\begin{eqnarray}
 \mathcal{L}(\{t,E, \vec x\}| \vartheta) &=& e^{-N_{astro} - N_{atm} - N_\mu}\prod_{i=1}^{N_{obs}} \bigg(
 N_{astro} P_{astro}(\vartheta) +  N_{atm} P_{atm} + N_{\mu} P_{\mu}
 \bigg),
 \label{eq:likedef}
\end{eqnarray}
where $N_{astro}, N_{atm}$, and $N_{\mu}$ are respectively the number of expected astrophysical neutrinos, atmospheric neutrinos, and atmospheric muons; these numbers are allowed to vary freely in our fits. The product runs over the observed events ($N_{obs} = 53$) in the dataset. The probability distributions of each component in Eq. \eqref{eq:likedef} are 
\begin{eqnarray}
P_{atm}(t_i,E_i,\vec x_i) &=& \sum_{f=e,\mu} \int dE_t d^3\vec x_{t}~ R_\Omega(\vec x_i,\vec x_{t}) R_E(E_i,E_t) A_{eff}(f,E_t,t_i,\vec x_{t}) P_{veto}(f,E_t,t,\vec x) \phi_{atm}(E_t,\vec x_{t}), \nonumber \\
P_{astro}(t_i,E_i,\vec x_i) &=& \sum_{f=e,\mu,\tau } \int dE_t d^3\vec x_{t}~ R_\Omega(\vec x_i,\vec x_{t}) R_E(E_i,E) A_{eff}(f,E_t,t_i,\vec x_{t}) \phi_{astro}(E_t,\vec x_{t}| \vartheta), \label{eq:Pastro}
\end{eqnarray}
and for $P_{\mu}(t_i,E_i,\vec x_i)$, we use the empirical parametrization given in Ref. \cite{Vincent:2016nut}.
We take into account the angular uncertainties of the events in $R_t(\vec x_i,\vec x_{t})$ by considering a 2d Gaussian point-spread-function centered around $\vec x_i$. Similarly we introduce the detector energy resolution assuming normal distributions. The probability of either track or cascade topology, and their mis-identification are also included in the likelihood.
Finally, $\phi(E,b,l)$ is the solution to the propagation equation \eqref{eqcascade}, where the galactic latitude and longitude $(b,l)$ implicitly specify DM column density. The model dependence of Eq. \eqref{eq:likedef} thus comes from the directional attenuation with respect to the isotropic hypothesis, $\mathcal{L} \propto \phi(E,b,l|\vartheta)/\phi^{iso}(E)$. 
\section{Constraints from the IceCube data and comparison with cosmological bounds}\label{results}
\begin{figure}[t]
\centering
\includegraphics[width=0.8\textwidth]{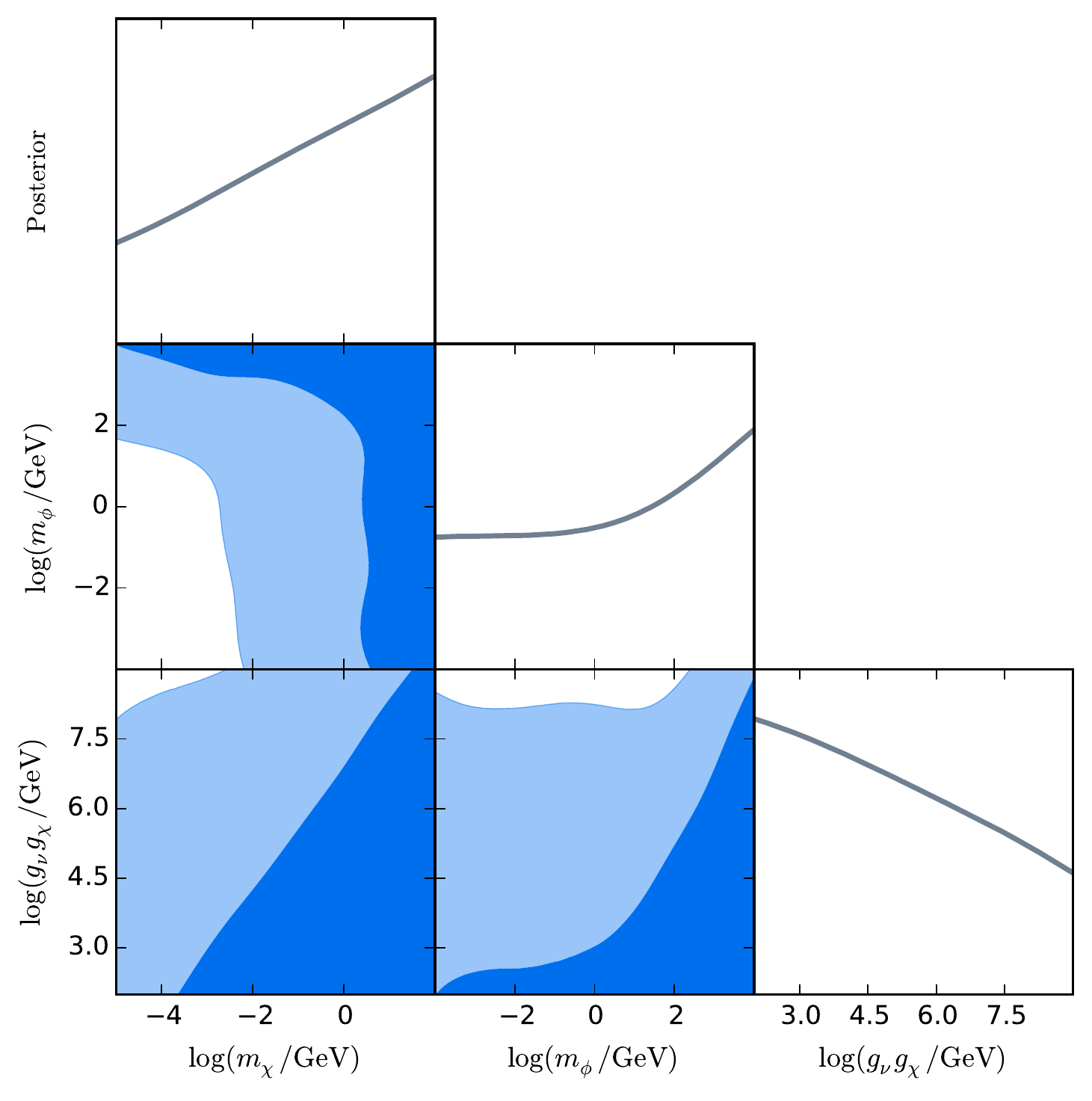}
\caption{One and two-dimensional posterior distributions in the space of dark matter mass $m_\chi$, mediator mass $m_\phi$ and couplings $g_\nu g_\chi$, as defined in Fig. 1. Dark and light blue regions specify the 68\% and 95\% containment regions.}
\label{fig:sspost}
\end{figure}

We produce constraints on the DM-$\nu$ scattering rate by evaluating Eq.~\eqref{eq:likedef} with the publicly available \texttt{emcee} \cite{2013PASP..125..306F} Markov Chain Monte Carlo software. We scan over with uniform priors in the space of \{$\log m_\chi$, $\log m_\phi$, $\log \left(g_\nu g_\chi\right)$, $N_{astro}$, $N_{atm}$, $N_\mu$\} and find the upper limits on the strength of the interaction in the parameter space of DM and mediator mass.
In Fig. \ref{fig:sspost} we show the posterior distributions as a function of the likelihood parameters. The split between astrophysical and atmospheric background events is consistent with previous results (see e.g. \cite{Vincent:2016nut}) and shows no correlation with the DM-neutrino interaction parameters.
These results can be compared with cosmological limits. At low energies, the DM-neutrino cross section in this case goes as $\sigma = g_\nu^2g_\chi^2 E_\nu^2 / 16\pi m_\phi^4m_\chi^4$. And the limit from the CMB for a cross section proportional to $E^2$ at low energies is \cite{Escudero:2015yka}: $\sigma < 10^{-40} (T/T_0)^2 (m_\chi/\mathrm{GeV})$ cm$^2$. Taking the average neutrino energy in the early universe and assuming a Fermi-Dirac distribution, this yields a limit on the couplings from cosmology
\begin{equation}
g_\chi g_\nu \lesssim 1.5\times 10^7 \left(\frac{m_\phi^4 m_\chi^3}{\mathrm{GeV}^7}\right)^{1/2} \, \mathrm{GeV}.
\label{eq:cosmo_limit}
\end{equation}
By inspection of Fig. \ref{fig:sspost} and using Eq. \eqref{eq:cosmo_limit}, one finds that with only 53 observed events IceCube constraints can rival cosmological constraints for $m_\phi \sim m_\chi \sim $ GeV, suggesting that future detectors such as IceCube Gen2 and KM3net will be able to significantly outperform CMB and LSS bounds.

\section{Acknowledgments}
We would like to thank Janet Conrad and Francis Halzen for useful comments and discussions. We would also like to thank Jordi Salvado Serra, Markus Ahlers, and  Rachel Carr. A.K. thanks the ICHEP 2016 organizers for the opportunity of presenting this work. C.A. is supported by NSF grants No. 1505858 and 1505855.  A.K. was supported in part by the U.S. NSF under Grants No. ANT-0937462 and PHY-1306958 and by the University of Wisconsin Research Committee with funds granted by the Wisconsin Alumni Research Foundation. A.C.V. is supported by an Imperial College Junior Research Fellowship and wishes to thank Francis Halzen for his gracious hospitality during his visit to WIPAC.

\bibliographystyle{JHEP}
\bibliography{DarkIce}
\end{document}